\documentstyle[12pt,epsfig]{article}
\begin{document}


\title{\quad  
{\Large{\bf Significance Calculation and a New Analysis Method\\ 
            in Searching for New Physics at the LHC}}
\author{ Yongsheng Gao, Liang Lu, Xinlei Wang\\
         Southern Methodist University\\
         Dallas, TX 75275-0175, USA } }
\maketitle

\begin{abstract}

The LHC experiments have great potential in discovering many possible 
new particles up to the TeV scale.
The significance calculation of an observation of a 
physics signal with known location and shape is no longer valid when 
either the location or the shape of the signal is unknown. 
We find the current LHC significance calculation of new physics 
is over-estimated and strongly depends on the 
specifics of the method and the situation it applies to.
We describe general procedures for significance calculation and 
comparing different search schemes.
A new method uses maximum likelihood fits with floating parameters 
and scans the parameter space for the best fit to the entire sample.
We find that the new method is significantly more sensitive than current 
method and is insensitive to the exact location 
of the new physics signal we search.

\end{abstract}


\section{Introduction}

The Large Hadron Collider (LHC) at CERN  
will open a new frontier in particle physics due to its higher 
collision energy and luminosity as compared to the existing accelerators. 
The general-purpose ATLAS and CMS experiments at the LHC will employ 
precision tracking, calorimetry and muon measurements over a large 
solid angle to identify and measure electrons, muons, photons, jets 
and missing energy accurately. Therefore, they have great physics potential 
in discovering many possible new particles. Among them are the Standard 
Model (SM) Higgs boson, supersymmetric (SUSY) and other new particles 
beyond the SM. All of them can have masses in a very large range up 
to the TeV scale.
The significance calculation in searching for and observation of a 
physics signal with known location and shape is no longer valid when 
either the location or the shape of the signal is unknown. 
This will be the case for many of the possible new physics signals at 
the LHC.

In Section 2, we give a short review of the significance calculation and 
current analysis strategy in High Energy Physics (HEP) and at the LHC.
Using a signal with known shape but unknown location as an 
example, we discuss in detail in Section 3 the problems of the current 
significance calculation.
We then describe general procedures for significance calculation and
comparing different search schemes in Section 4.
In Section 5, we describe a new analysis method and compare it
with the current ``Sliding-Window'' approaches following these procedures.
Detailed comparison results are also given in this Section. 
Summary and discussion are given in Section 6. 
In this note we limit ourselves to the significance calculation and 
analysis method used in searching for an individual decay mode of new 
physics signals.

\section{Review of Significance Calculation and Current Analysis Strategy 
         to Search for New Physics at the LHC}

In the field of HEP, a common strategy to detect a physics signal
is to search for an excess of events in a certain region of
a kinematic observable. The observation probability is given by Poisson 
statistics:

\begin{equation}
P(n,B)=\frac{e^{-B} B^n}{n!}
\label{poisson1}
\end{equation}
where $B$ is the number of the expected events to be observed in the region, 
and $n$ is the number of the observed events
in this region. When $B$ is large (over 25, for instance), the significance 
of an observation can be approximated well by $S/\sqrt{B}$ of Gaussian 
statistics, where $S=n-B$. 

In HEP, the significance of an observation is defined by the probability 
that such an observation is due to statistical fluctuation of background
events.
When we claim an observation has a significance of $5\sigma$~\cite{PDG},
the common criterion for a HEP discovery, the probability that the claimed 
discovery 
is due to statistical fluctuation of background events, known as the Type 
I error rate in statistics, needs to be less than $2.9\times 10^{-7}$. 
The background fluctuation probabilities which define the 1$\sigma$ to 
5$\sigma$ significances in HEP are shown in Table~\ref{tab:sig-def}.

\begin{table}[hhh]
\caption{The definition of significance in HEP and the corresponding
         background statistical fluctuation probabilities.}
\begin{center}
\begin{tabular}{|c|c|c|c|c|c|} 
\hline
 Significance & $1 \sigma$ &$2 \sigma$ &$3 \sigma$ &$4 \sigma$ &$5
 \sigma$                                                            \\ \hline
Probability that the observation  & & & & & \\
of the excess of events is due to & 15.87\% & 2.28\% & 0.14\% & 3.2$\times 10^{-5}$ 
                                                  & 2.9$\times 10^{-7}$      \\
background statistical fluctuation   & & & & &                              \\ \hline
\end{tabular}
\end{center} 
\label{tab:sig-def}
\end{table}

If the expected mass spectrum of a
physics signal is a Gaussian distribution with standard deviation $\sigma$, 
the mass region used to calculate the observation significance of this 
signal is usually $\pm 2 \sigma$ around the Gaussian mean. 
Including regions where the physics signal has little chance to
show up only increases $B$ and decreases $S/\sqrt{B}$.  
This is why the region in which to search for the signal and calculate 
significance is 
usually limited to $\pm 2 \sigma$ around the Gaussian mean, in order 
to maximize the discovery potential and observation significance.
This approach has been widely and successfully used in many HEP experiments 
at CESR, Tevatron, LEP, KEK-B, PEP-II, etc. It is only valid when searching 
for and observation of a physics signal with known location and shape, i.e., 
when the kinematic region for the significance calculation is uniquely 
defined.

One of the new challenges for the ATLAS and CMS experiments is that we do not 
know the masses of the new particles we will be searching for. 
The current analysis method proposed for new particle searches at the LHC is 
to use a ``Sliding-Window'', i.e., look for an excess of events in a series of 
narrow regions or windows over the entire available kinematic range. The location 
and width of each window is given by the expectations of the new particle
with a specific mass and the corresponding width. 
The expected significances and discovery potential for new particle searches 
are only determined by the $S$ and $B$ values within these narrow
windows~\cite{atlas-tdr-15,atlasnotes,cmsnotes}.

\section {Problem with the Current LHC Search Method and Significance Calculation}

There is a fundamental problem in the above significance calculation.
The significance of an observation is defined according to the 
probability that such an observation is due to statistical fluctuation 
of background events, i.e., the Type I error rate. 
The current expected significance calculation is only correct 
if we know exactly the location and shape of the new physics signal 
we are searching for, and we use only one window to search and 
calculate the observation significance.
In the ``Sliding-Window'' method, we search for an excess of 
events in any of the narrow windows over a wide kinematic range, 
but still use the $S$ and $B$ of each narrow window 
to calculate the significance of the
observation. Therefore, the probability of observing a ``significant'' 
excess of events due to background statistical fluctuation in any 
window will be much higher~\cite{hepex-0310011}.
This ``false-positives'' problem caused by multiple testing was 
recognized in statistics many years ago~\cite{Lehmann,Neill}.

We use simple simulations to demonstrate this problem. Assume that
we search for a possible Gaussian signal with a standard 
derivation $\sigma$=1.0 but an unknown mean between 2.0 
and 98.0, and that the expected distribution of the background 
is flat between 0.0 and 100.0. 
We generate 13,450,000 background-only Monte Carlo (MC) experiments
(referred to as the ``background-only sample'') 
with each experiment containing 500 events generated from a flat 
distribution between 0.0 and 100.0.

We use a ``Sliding-Window'' with a fixed width of 4.0 and move the center 
of this fixed-width window from 2.0 to 98.0 with various step sizes of 
16.0, 8.0, 4.0, 2.0, 1.0, 0.5, 0.2 and 0.1, respectively to search for
an excess of events in any of the windows. The fixed  width of 4.0 of 
the ``Sliding-Window'' corresponds to $\pm 2 \sigma$ around the unknown 
Gaussian mean.
The significance in any one window of a MC experiment is calculated by
$S/\sqrt{B}$ according to the current significance calculation, 
where $n$ is the number of events in that window of the experiment,
$S=n-B$, and $B$ = 20.

The probabilities that we observe at least one window with $S/\sqrt{B}$
$>$ 1, 2, 3, 4, 5 (i.e. ``1'', ``2'', ``3'', ``4'' and ``5''$\sigma$
according to the current significance calculation) in any of the
background-only sample experiment are shown in Table~\ref{tab:table2}.
This probability is defined as the number of background-only MC 
experiments which contain at least one ``1'', ``2'', ``3'', ``4'' 
and ``5''$\sigma$-effect window divided by the total number of 
background-only MC experiments.
From Table~\ref{tab:table2}, we can see that the probabilities of 
positive observations are much higher than the Table~\ref{tab:sig-def}
background fluctuation probabilities that define the significances in HEP.
Furthermore, the probability of finding a signal of given 
significance increases as the step size of the 
``Sliding-Window'' decreases, i.e., as more windows are scanned over the 
same kinematic range. 
While each individual window follows Poisson or Gaussian statistics
reasonably well, 
the probability of observing an excess in any of the multiple windows 
is much higher than that for an individual window.
Table~\ref{tab:table2} clearly shows the problems of the 
significance calculation in searching for new physics signals with 
an unknown location. It is due to the fact that
we search for an excess of events 
over multiple narrow windows, but the significance is still calculated 
according to an individual narrow window.

\begin{table}[hhh]
\caption{The probability of observing at least one ``1'', ``2'', 
         ``3'', ``4'' and ``5''$\sigma$-effect window in any background-only 
         MC experiment using the ``Sliding-Window'' method with various 
         step sizes.}
\begin{center}
\begin{tabular}{|c|c|c|c|c|c|}   \hline
Significance ($S/\sqrt{B}$)   
                  &  ``1''$\sigma$ & ``2''$\sigma$ & ``3''$\sigma$ 
                  &  ``4''$\sigma$ & ``5''$\sigma$          \\ \hline
Step Size = 16    & 70.89\% & 20.42\% & 1.522\%
                  & 0.11\%  & 0.002\%                       \\ \hline
Step Size = 8     & 91.56\% & 35.25\% & 2.818\%
                  & 0.20\%  & 0.003\%                       \\ \hline
Step Size = 4     & 99.72\% & 58.53\% & 5.380\%
                  & 0.39\%  & 0.007\%                       \\ \hline
Step Size = 2     & 99.99\% & 77.86\% & 9.635\%
                  & 0.73\%  & 0.015\%                       \\ \hline
Step Size = 1     & 100.0\% & 89.03\% & 14.86\%
                  & 1.24\%  & 0.027\%                       \\ \hline
Step Size = 0.5   & 100.0\% & 94.33\% & 19.97\%
                  & 1.83\%  & 0.042\%                       \\ \hline
Step Size = 0.2   & 100.0\% & 97.17\% & 25.42\%
                  & 2.56\%  & 0.064\%                       \\ \hline
Step Size = 0.1   & 100.0\% & 98.01\% & 28.21\%
                  & 2.98\%  & 0.078\%                       \\ \hline 
\end{tabular} 
\end{center}
\label{tab:table2}
\end{table}

\vskip 0.3cm

\section{Procedures for Significance Calculation and Comparing 
         Different Analysis Approaches.}

Each analysis approach or search scheme can be described by two measures.
The Type I error rate measures how often false signals are claimed 
when there are only background events. The significance of an 
observation as defined according to this error rate is 
shown in Table~\ref{tab:sig-def}.
The Power or Sensitivity measures 
how often real signals can be found correctly when they are present. 
There is a correlation between the two measures, the Type I error rate 
increases with increasing sensitivity. Therefore, we need to set one 
of these two measures to the same value for different search schemes 
and compare the other measure, in order to quantitatively 
compare these search schemes.

We can see from Table~\ref{tab:table2} that the ``significance'' 
calculated by $S/\sqrt{B}$ of the ``Sliding-Window'' method is highly 
over-estimated compared to the HEP significance definition. 
Furthermore, it strongly depends on the specifics of the search scheme 
and the situation it applies to, i.e. step size of the ``Sliding-Window'' 
used to scan the kinematic range, the total range of the kinematic region, 
etc.
We need to evaluate the significance reported by each
scheme so it truly reflects our significance definition. 
The procedures to calculate significance and compare different search 
schemes are as follows:

\begin{enumerate}

\item{Use background-only MC experiments to evaluate the significance
      of all search schemes. After the evaluation, all the search schemes
      should be normalized to have the same Type I error rates, 
      which follow the HEP significance definition.}

\item{Use signal-embedded MC experiments to evaluate the  
      sensitivity of the search schemes. The search scheme with the higher 
      sensitivity is the better one.}

\end{enumerate}

These procedures are applied to compare a new analysis method with the 
current ``Sliding-Window''approaches in the following Section.

\section{A New Analysis Method and a Comparison 
            with the ``Sliding-Window'' Approaches}

An alternative approach is to apply an unbinned maximum 
likelihood scan method with floating parameters to the entire sample 
and search for the best fit to the sample over the entire parameter 
space~\cite{hepex-0310011}. 
It is intended to minimize the sensitivity of the significance to local 
fluctuations.
We follow the procedures described in Section 4 to compare the current 
``Sliding-Window'' approaches with this new method for this 
example~\cite{liangthesis}.

\begin{enumerate}

\item{We search for a possible Gaussian signal ($\sigma$=1.0 with unknown 
      mean between 2.0 and 98.0) on top of a flat background in the
      13,450,000 background-only MC experiments (``background-only sample'')
      using each search scheme. 
      We then evaluate the significance of each scheme so that it follows the 
      HEP significance definition for the background-only sample.}

\item{We generate signal-embedded MC experiments and perform the
      same search using each search 
      scheme. We then calculate the sensitivities of finding the embedded 
      signal for each search scheme based on the significances 
      defined by the background-only sample.}

\end{enumerate}

For the ``Sliding-Window'' approach in Step 1, we make a table which 
defines the new cutoff values of $S/\sqrt{B}$ which follow the HEP 
significance definition for the background-only sample. 
Similarly for the new approach, we find out the 
values of the Maximum Likelihood fit output which corresponds to 1, 
2, 3, 4 and 5$\sigma$ for the background-only sample
according to the HEP significance definition.

\subsection{Significance Evaluation of ``Sliding-Window'' Approaches}

\begin{figure}[htbp]
\epsfxsize=0.90\textwidth\centerline{\mbox{\epsffile{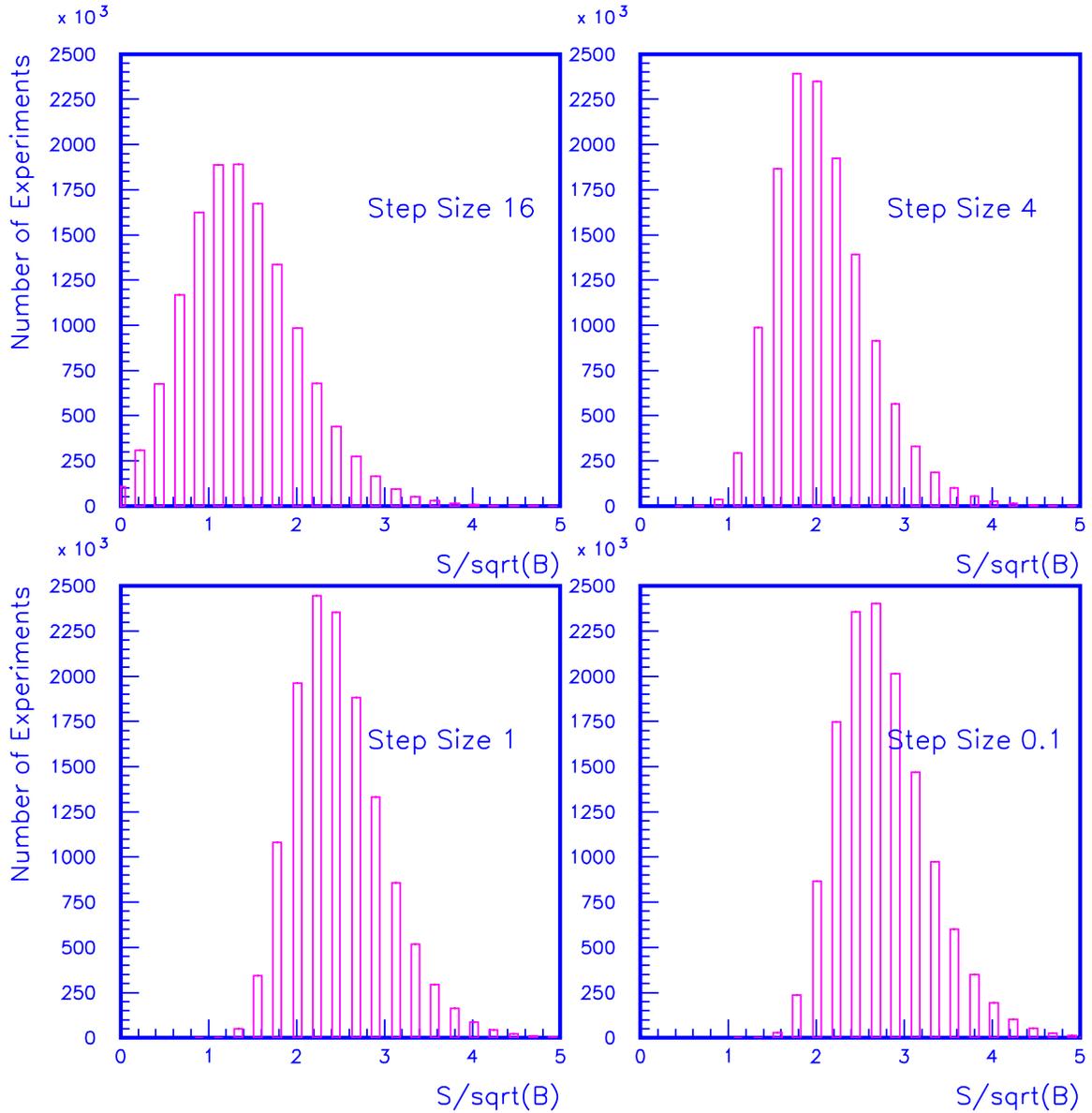}}}
\caption{The maximum  $S/\sqrt{B}$ from ``Sliding-Window'' approaches
         with step sizes of 16, 4, 1, and 0.1 for the 13,450,000 
         background-only MC experiments.}
\label{fig:SW-bkg}
\end{figure}

We use the background-only sample to evaluate the 
significance of the ``Sliding-Window'' approach. 
For each experiment, we use a ``Sliding-Window'' with fixed width 
of 4 and move the center of the window from 2.0 to 98.0 with step sizes 
of 16, 8, 4, 2, 1, 0.5, 0.2 and 0.1, respectively, to search for the 
window with the maximum $S/\sqrt{B}$.
For each step size, we plot the maximum $S/\sqrt{B}$ for all the 
background-only sample. We then find the corresponding 
cutoff values on the plot which follow the HEP significance definition.
For example, the maximum  $S/\sqrt{B}$ from ``Sliding-Window'' approaches
with step sizes of 16, 4, 1, and 0.1 for the background-only sample
are shown in Figure~\ref{fig:SW-bkg}.
In the ``Sliding-Window'' approach with step size of 0.1,
we find that 15.87\% of the experiments have at least one window with
$S/\sqrt{B}$  $>$ 3.35, and 2.28\% of the experiments have at least one
window with $S/\sqrt{B}$  $>$ 4.02. According to our HEP significance 
definition in Table~\ref{tab:sig-def}, the experiments which contain 
windows with $S/\sqrt{B}$ $>$ 3.35 are defined as 1$\sigma$ for the 
``Sliding-Window'' approach with step size of 0.1 in this case.
Similarly, the experiments which contain window with $S/\sqrt{B}$ 
$>$ 4.02 are defined as 2$\sigma$.
The new $S/\sqrt{B}$ cutoff values which follow our HEP significances 
definition for the ``Sliding-Window'' approaches with various step sizes 
are given in Table~\ref{tab:SW-rescale}.
The cutoff values are not continuous, because $S=n-B$, $B=20$, and both
$n$ and $S$ are integers.

\begin{table}[hhh]
\caption{The $S/\sqrt{B}$ cutoff values which correspond to the HEP 
         significance definition for a ``Sliding-Window'' approach
         with different step sizes.}
\begin{center}
\begin{tabular}{|c|c|c|c|c|c|} 
\hline
Significance    & 1$\sigma$ & 2$\sigma$ & 3$\sigma$ & 4$\sigma$ 
                & 5$\sigma$  \\ \hline
Step Size =  16 & 2.01 & 2.90 & 3.80 & 4.91 & 6.03 \\ \hline
Step Size =   8 & 2.23 & 3.13 & 4.02 & 5.14 & 6.26 \\ \hline
Step Size =   4 & 2.68 & 3.35 & 4.24 & 5.36 & 6.48 \\ \hline
Step Size =   2 & 2.90 & 3.57 & 4.47 & 5.36 & 6.48 \\ \hline
Step Size =   1 & 2.90 & 3.80 & 4.47 & 5.59 & 6.48 \\ \hline
Step Size = 0.5 & 3.13 & 3.80 & 4.69 & 5.81 & 6.93 \\ \hline
Step Size = 0.2 & 3.13 & 4.02 & 4.91 & 5.81 & 6.93 \\ \hline
Step Size = 0.1 & 3.35 & 4.02 & 4.91 & 5.81 & 6.93 \\ \hline
\end{tabular}
\end{center} 
\label{tab:SW-rescale}
\end{table}

\subsection{Significance Evaluation of the New Analysis Method}

We use the same background-only sample to 
evaluate the significance for the new approach. 
In this specific example we search for a Gaussian signal ($\sigma$=1.0 
with unknown mean between 2.0 and 98.0) on top of a uniform background. 
The Likelihood is then calculated as:

\begin{equation}
L(Y|\mu) = \prod_{i=1}^{n} P(y_i|\mu)
\end{equation}
where the Y are the data per experiment, and $y_{i}$ is the individual 
data point in each experiment where i = 1, 2, 3, ..., n (n = 500).
$P(y_{i}|\mu)$ is the normalized probability density of $y_{i}$ as a function of 
the parameter $\mu$ which is the unknown mean of the Gaussian signal.
The normalized probability density is given by:

\begin{equation}
 P(y_i|\mu) = \frac{(1-p)}{100} + \frac{p}{\sqrt{2\pi}} e^{-\frac{1}{2}(y_i-\mu)^2} 
\end{equation}
where 100 is the normalization factor which guarantees that the integral of 
$P(y_i|\mu)$ over the range from 0.0 to 100.0 is equal to 1. 
$p$ is the probability of the data point being the Gaussian signal.
Similarly, ($1-p$) is the probability of the data point being the background.
The optimization process attempts to find the $\mu$ parameter that maximizes
$L(Y|\mu)$ for each experiment, or minimizes $-log(L(Y|\mu))$.
We use  

\begin{equation}
\sum_{i=1}^{500} log((1-p)+ p \frac{100}{\sqrt{2\pi}}e^{-\frac{1}{2}
(y_i-\mu)^2})
\end{equation}
as the maximum likelihood output
to simplify the calculation.

For the MC experiment generation and maximum likelihood analysis we used the 
statistical computing software R~\cite{R,R2}. R is a language and environment 
for statistical computing and graphics. It is a GNU project
developed at Bell Laboratories and provides a wide variety of
statistical and graphical techniques (linear and nonlinear modeling, 
classical statistical tests, time-series analysis, classification, 
clustering, etc.)~\cite{R3}. 
We have also tried other statistical software packages such as
SAS~\cite{SAS}, Matlab\cite{Matlab} and the HEP software package
RooFit~\cite{Roofit} to generate the MC experiments, and perform the
maximum likelihood fits.
The results with different analysis tools are all consistent. 
We decided to use R because it is faster than the other packages.

\begin{figure}[htbp]
\epsfxsize=0.90\textwidth\centerline{\mbox{\epsffile{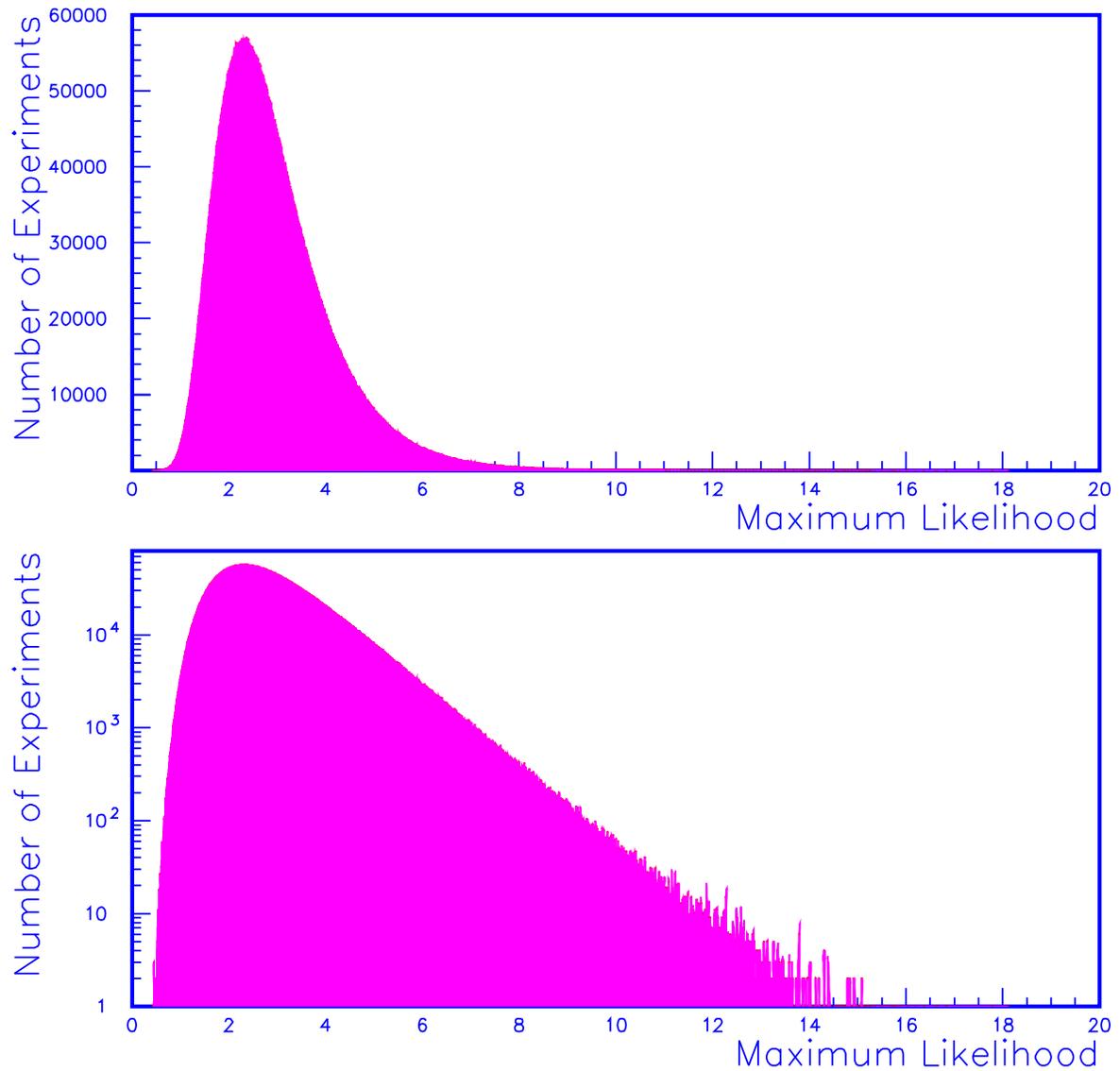}}}
\caption{The Maximum Likelihood output for 13,450,000 background-only
         MC experiments. The top plot is in linear scale while the bottom 
         plot is in log scale.}
\label{fig:ML-BKG}
\end{figure}

In order to find the best fit with a floating $\mu$ parameter for 
each experiment,
we break down the $\mu$ parameter region from 2.0 to 98.0 into 
96 equal intervals~\cite{kirkby}. 
We perform one maximum likelihood fit for each interval to 
find the best fit. We then compare all 96 fits to find 
the overall best fit for the entire $\mu$ parameter space 
for this experiment.
The maximum likelihood output of the best fit for the background-only
sample is shown in Figure~\ref{fig:ML-BKG}.  
Because 84.13\% of the background-only MC experiments have a maximum
likelihood output below 4.00, the cutoff value for 1$\sigma$ is set at 4.00. 
Similarly, 5.94 is set as the cutoff value for 2$\sigma$. The cutoff  values 
for 1 to 5$\sigma$ significances for the new analysis method in this 
example are given in Table~\ref{tab:ML-rescale}. 

\begin{table}[hhh]
\caption{The cutoff values of the maximum likelihood output for
         the new analysis method in this example.}
\begin{center}
\begin{tabular}{|c|c|c|} 
\hline
Fraction of background-only    & Cutoff value & Significance   \\ 
experiments below cutoff       &              &                \\ \hline
    84.13\%                    &  4.00   & 1$\sigma$           \\ \hline
    97.72\%                    &  5.94   & 2$\sigma$           \\ \hline
    99.86\%                    &  8.71   & 3$\sigma$           \\ \hline
    99.9968\%                  & 12.48   & 4$\sigma$           \\ \hline
    $(1-2.9\times 10^{-5})$\%  & 16.61   & 5$\sigma$           \\ \hline
\end{tabular}
\end{center} 
\label{tab:ML-rescale}
\end{table}

After these evaluations, the significances reported by the 
``Sliding-Window'' approaches and the new analysis method are all
adjusted to follow the HEP 
significance definition for background-only sample. 
We can then compare the sensitivity of these approaches using the
signal-embedded MC experiments.

\subsection{Sensitivity Comparison of ``Sliding-Window'' Approaches 
            with the New Analysis Method}

We generate signal-embedded MC experiments to calculate the Power or 
Sensitivity of each approach. Each experiment contains a small 
number (5, 10, 15, 20, 25, 30, and 35) of signal events generated 
according to a
Gaussian distribution with $\sigma$=1.0 and a specific 
Gaussian mean (42.00, 46.00, 48.00, 49.00, 49.50, 49.75, 49.90, 49.95,
and 50.00)~\cite{liangthesis}. 
Each signal-embedded experiment contains one set of these 
signal events embedded with 500 background events generated with a flat 
distribution between 0.0 and 100.0. 
A total of 630,000 signal-embedded MC experiments are generated, with
10,000 experiments for each set of Gaussian signal parameters.
For example, 10,000 experiments each with 5 Gaussian signal events 
with Gaussian mean at 42.00 embedded into 500 background events
are generated.
The maximum likelihood output for each signal-embedded experiment is 
normalized to 500 events before comparing with the cutoff values from 
background-only experiments.

We use these signal-embedded experiments to calculate the sensitivity 
of the ``Sliding-Window'' approaches.
We use a ``Sliding-Window'' of fixed width of 4.0 and move the center 
of this fixed-width window from 2.0 to 98.0 with various step sizes of 
16.0, 8.0, 4.0, 2.0, 1.0, 0.5, 0.2 and 0.1, respectively, to search for
the window with the maximum $S/\sqrt{B}$ for each experiment.
The success of finding the embedded signal in an experiment 
is defined as when the center of the most significant window is found 
within 1.0 of the Gaussian mean of the embedded 
signal events. The significance of this window is defined according to the 
cutoff values in Table~\ref{tab:SW-rescale}.

We use the same signal-embedded experiments to calculate the sensitivity 
of the new analysis method. 
For each experiment, we break down the $\mu$ parameter region from 2.0 to 
98.0 into 96 equal intervals. We perform one maximum likelihood fit for each 
interval to find the best fit for this interval and the corresponding $\mu$ 
value. 
We then compare all 96 fits to find the best overall fit for the entire $\mu$ 
parameter space and its corresponding $\mu$ value for each 
experiment.
The success of finding the embedded signal in an experiment is defined as 
when the $\mu$ value of the best fit in the entire $\mu$ parameter space 
falls within 1.0 of the Gaussian mean of the embedded 
signal events. Similarly, the significance is defined according to the
cutoff values in Table~\ref{tab:ML-rescale}.

The sensitivity of the ``Sliding-Window'' approaches with various step sizes 
is very sensitive to the exact value of the Gaussian mean 
of the embedded signal events. Therefore, we choose the 
best-case and worst-case scenarios for each ``Sliding-Window'' approach 
with a specific step size. 
The best-case scenario corresponds to a case 
where the $\pm 2 \sigma$ region of the embedded Gaussian signal falls exactly 
inside one of the ``Sliding-Windows''.
The worst-case scenario corresponds to a case when the embedded 
Gaussian signal falls exactly between two neighboring ``Sliding-Windows''.
In Table~\ref{tab:best-worst}, we show the Gaussian means of the embedded 
Gaussian signals for the best-case and worst-case scenarios for various 
``Sliding-Window'' step sizes.

\begin{table}[hhh]
\caption{The Gaussian means of the embedded Gaussian signal
         for the best-case and worse-case scenarios of the 
         ``Sliding-Window'' approaches with various step sizes.}
\begin{center}
\begin{tabular}{|c|c|c|}\hline
Step Size  &  Gaussian Mean of embedded signal   
           &  Gaussian Mean of embedded signal              \\  
           &  Best-Case Scenario  
           &  Worst-Case Scenario                           \\ \hline 
  16       &         50.00     &       42.00                 \\ \hline
   8       &         50.00     &       46.00                 \\ \hline
   4       &         50.00     &       48.00                 \\ \hline
   2       &         50.00     &       49.00                 \\ \hline
   1       &         50.00     &       49.50                 \\ \hline
   0.5     &         50.00     &       49.75                 \\ \hline
   0.2     &         50.00     &       49.90                 \\ \hline
   0.1     &         50.00     &       49.95                 \\ \hline
\end{tabular} 
\end{center}
\label{tab:best-worst}
\end{table}

For each set of the 10,000 signal-embedded MC experiments, we calculate how
many times the embedded signals are correctly found by each approach 
with a significance
greater than 1, 2, 3, 4 and 5$\sigma$ according to the HEP significance
definitions. For the ``Sliding-Window'' approach with a specific step
size, two numbers are reported, according to the best-case and worst-case
scenarios respectively. 
In contrast, the new approach scans the parameter space and performs a 
maximum likelihood fit at each small interval to cover the entire parameter 
space to search for the best fit of the entire sample. Thus, it is not 
sensitive to the exact value of the Gaussian mean of the embedded signal.
We find the number is independent of the exact location of the embedded 
Gaussian signal for the new analysis method.

\begin{table}[hhh]
\caption{The number of signal embedded experiments successfully found with 1,
         2, 3, 4 and 5$\sigma$ 
         significance in the 10,000 MC experiments each with 10 signal 
         events embedded.}
\begin{center}
\begin{tabular}{|c|c|c|c|c|c|}
\hline
Significance & 1$\sigma$  & 2$\sigma$  & 3$\sigma$  & 4$\sigma$  & 5$\sigma$ \\
Scenario     & Best/Worst & Best/Worst & Best/Worst & Best/Worst & Best/Worst\\
\hline
Step Size = 16  & 4795/0    & 1859/0  & 433/0   & 39/0 & 1/0 \\
Step Size =  8  & 3897/0    & 947/0   & 271/0   & 23/0 & 1/0 \\
Step Size =  4  & 1839/0    & 652/0   & 176/0   & 16/0 & 1/0 \\
Step Size =  2  & 1691/0    & 619/0   & 107/0   & 16/0 & 1/0 \\
Step Size =  1  & 1915/1031 & 537/274 & 143/68  &  9/1 & 1/0 \\
Step Size = 0.5 & 2011/1728 & 807/706 & 156/120 & 11/2 & 2/0 \\
Step Size = 0.2 & 2011/1728 & 562/484 & 97/70   & 11/2 & 2/0 \\
Step Size = 0.1 & 1598/1465 & 599/548 & 106/78  & 13/2 & 1/0 \\ \hline
New Approach    & 2328      & 819     & 153     & 17   & 2   \\ \hline
\end{tabular} 
\end{center}
\label{result10}
\end{table}

\begin{table}[hhh]
\caption{The number of signal embedded experiments successfully found with 1,
         2, 3, 4 and 5$\sigma$ 
         significance in the 10,000 MC experiments each with 20 signal 
         events embedded.}
\begin{center}
\begin{tabular}{|c|c|c|c|c|c|}
\hline
Significance & 1$\sigma$  & 2$\sigma$  & 3$\sigma$  & 4$\sigma$  & 5$\sigma$\\
Scenario     & Best/Worst & Best/Worst & Best/Worst & Best/Worst & Best/Worst\\
\hline
Step Size = 16  & 9834/0    & 8969/0    & 6302/0    & 2215/0    & 356/0   \\
Step Size =  8  & 9694/0    & 7844/0    & 5435/0    & 1616/0    & 240/0   \\
Step Size =  4  & 8932/0    & 7135/0    & 4542/0    & 1190/0    & 158/0   \\
Step Size =  2  & 8439/0    & 6867/0    & 3642/0    & 1182/0    & 156/0   \\
Step Size =  1  & 8086/4955 & 6365/3913 & 4032/2509 & 965/634   & 174/112 \\
Step Size = 0.5 & 7795/7264 & 6812/6273 & 4020/3782 & 1046/965  & 130/100 \\
Step Size = 0.2 & 7795/7264 & 6217/5747 & 3274/3061 & 1046/965  & 130/100 \\
Step Size = 0.1 & 7538/7277 & 6314/6034 & 3416/3268 & 1119/1051 & 149/111 \\
\hline
New Approach    & 9116      & 7406      & 4073      & 1209      & 206 \\ \hline
\end{tabular} 
\end{center}
\label{result20}
\end{table}

\begin{table}[hhh]
\caption{The number of signal embedded experiments successfully found with 1,
         2, 3, 4 and 5$\sigma$ 
         significance in the 10,000 MC experiments each with 30 signal 
         events embedded.}
\begin{center}
\begin{tabular}{|c|c|c|c|c|c|}
\hline
Significance &  1$\sigma$ & 2$\sigma$  & 3$\sigma$  & 4$\sigma$  & 5$\sigma$ 
\\
Scenario     & Best/Worst & Best/Worst & Best/Worst & Best/Worst & Best/Worst\\
\hline
Step Size = 16  & 10000/0   & 10000/0   & 9963/0    & 9215/0    & 5936/0    \\
Step Size =  8  & 10000/0   & 9993/0    & 9917/0    & 8792/0    & 5079/0    \\
Step Size =  4  & 9998/0    & 9987/0    & 9841/0    & 8249/0    & 4238/0    \\
Step Size =  2  & 9904/0    & 9886/0    & 9633/0    & 8210/0    & 4232/0    \\
Step Size =  1  & 9542/6589 & 9532/6584 & 9412/6490 & 7787/5444 & 4560/3285 \\
Step Size = 0.5 & 9249/8784 & 9246/8781 & 9152/8665 & 7743/7331 & 3847/3683 \\
Step Size = 0.2 & 9249/8784 & 9238/8778 & 9046/8558 & 7743/7331 & 3847/3683 \\
Step Size = 0.1 & 9183/8945 & 9178/8934 & 9006/8742 & 7792/7603 & 4049/3944 \\
\hline
New Approach    & 9985      & 9974      & 9723      & 8024      & 4332 \\ \hline
\end{tabular} 
\end{center}
\label{result30}
\end{table}

\begin{figure}[htbp]
\epsfxsize=0.90\textwidth\centerline{\mbox{\epsffile{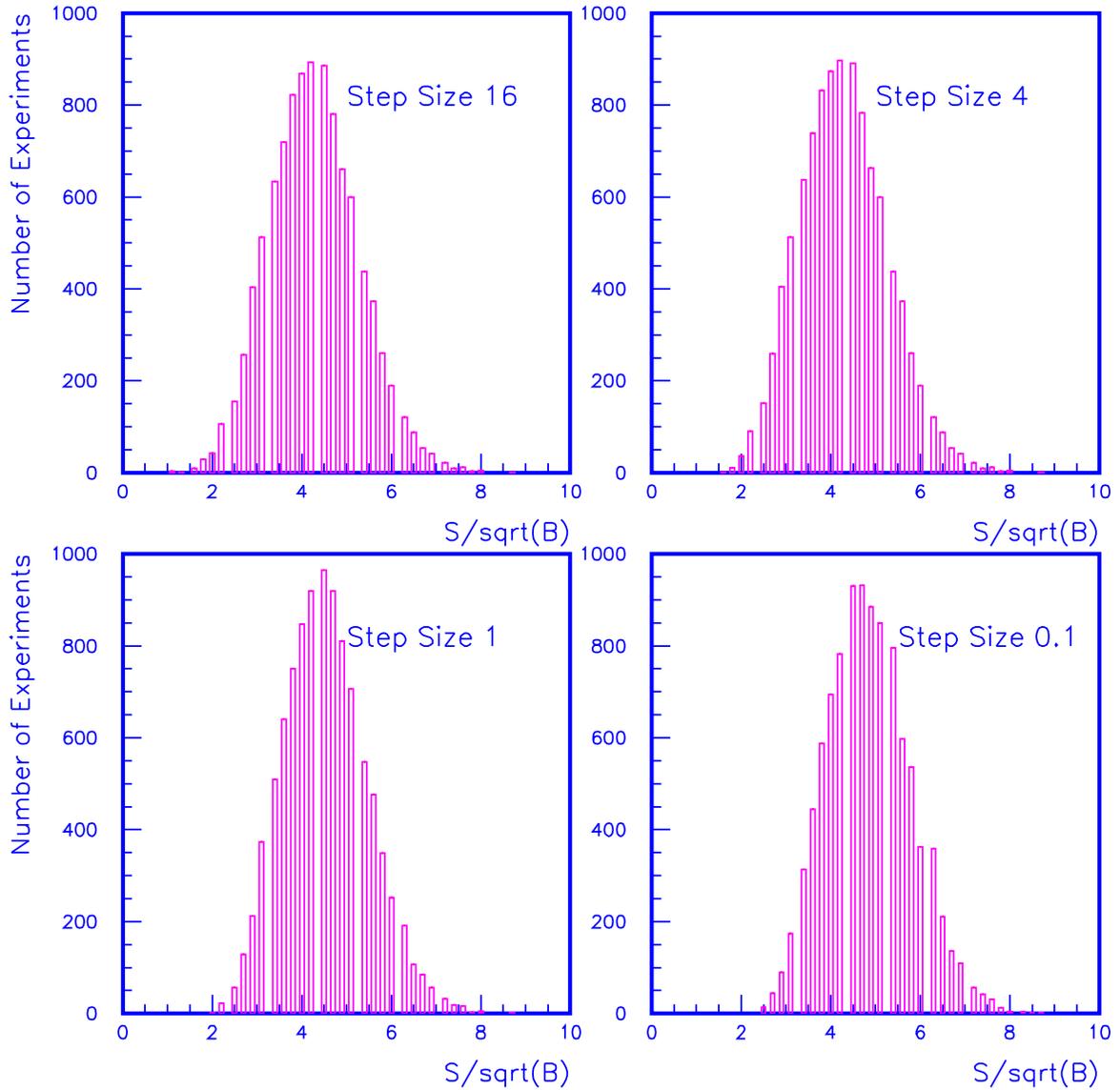}}}
\caption{The maximum  $S/\sqrt{B}$ from ``Sliding-Window'' approaches
         with step sizes of 16, 4, 1, and 0.1 for the best-case scenario
         10,000 MC experiments each with 20 signal events embedded.}
\label{fig:SW-20-best}
\end{figure}

\begin{figure}[htbp]
\epsfxsize=0.90\textwidth\centerline{\mbox{\epsffile{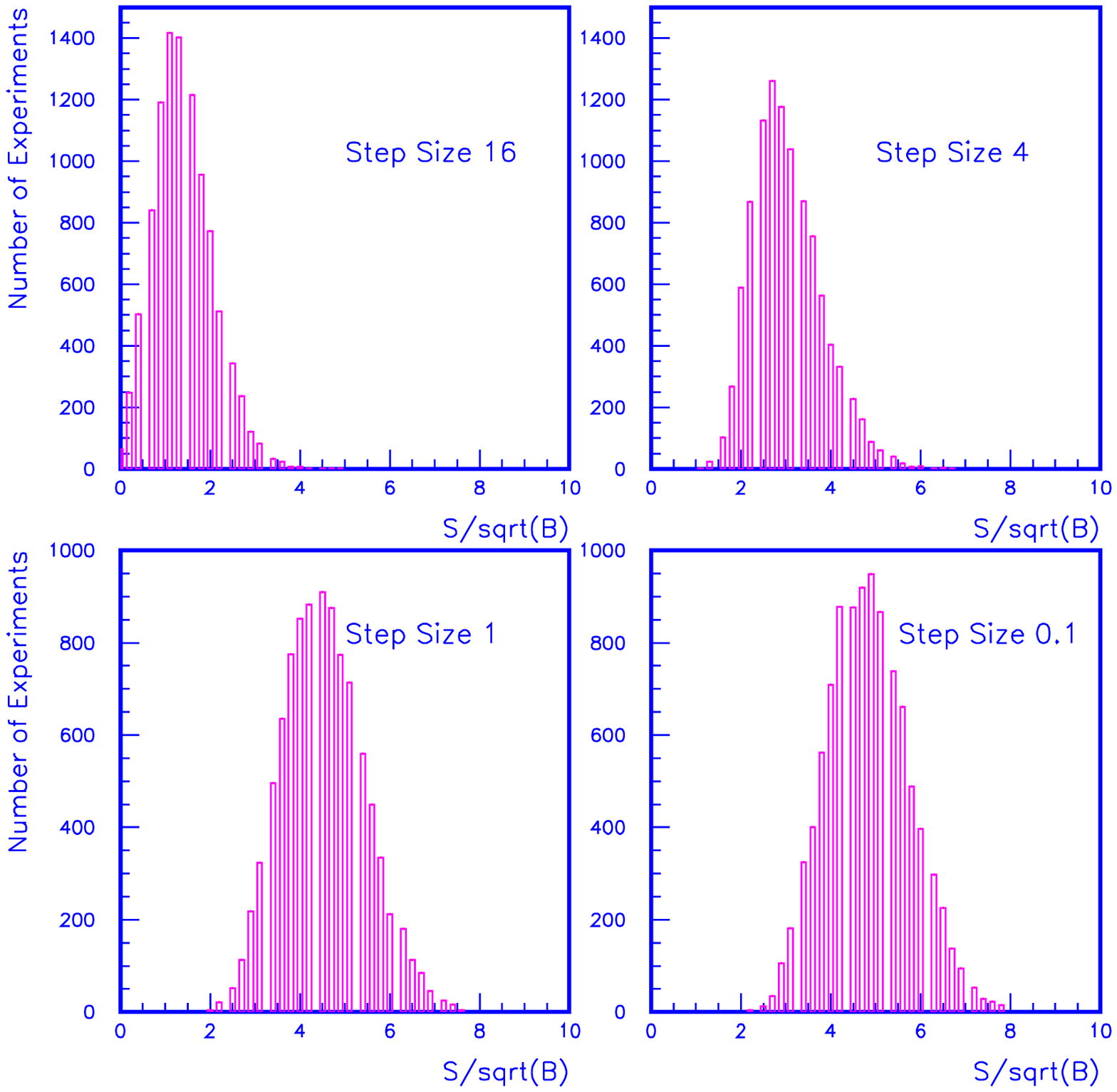}}}
\caption{The maximum  $S/\sqrt{B}$ from ``Sliding-Window'' approaches
         with step sizes of 16, 4, 1, and 0.1 for the worst-case scenario
         10,000 MC experiments each with 20 signal events embedded.}
\label{fig:SW-20-worst}
\end{figure}

\begin{figure}[htbp]
\epsfxsize=0.90\textwidth\centerline{\mbox{\epsffile{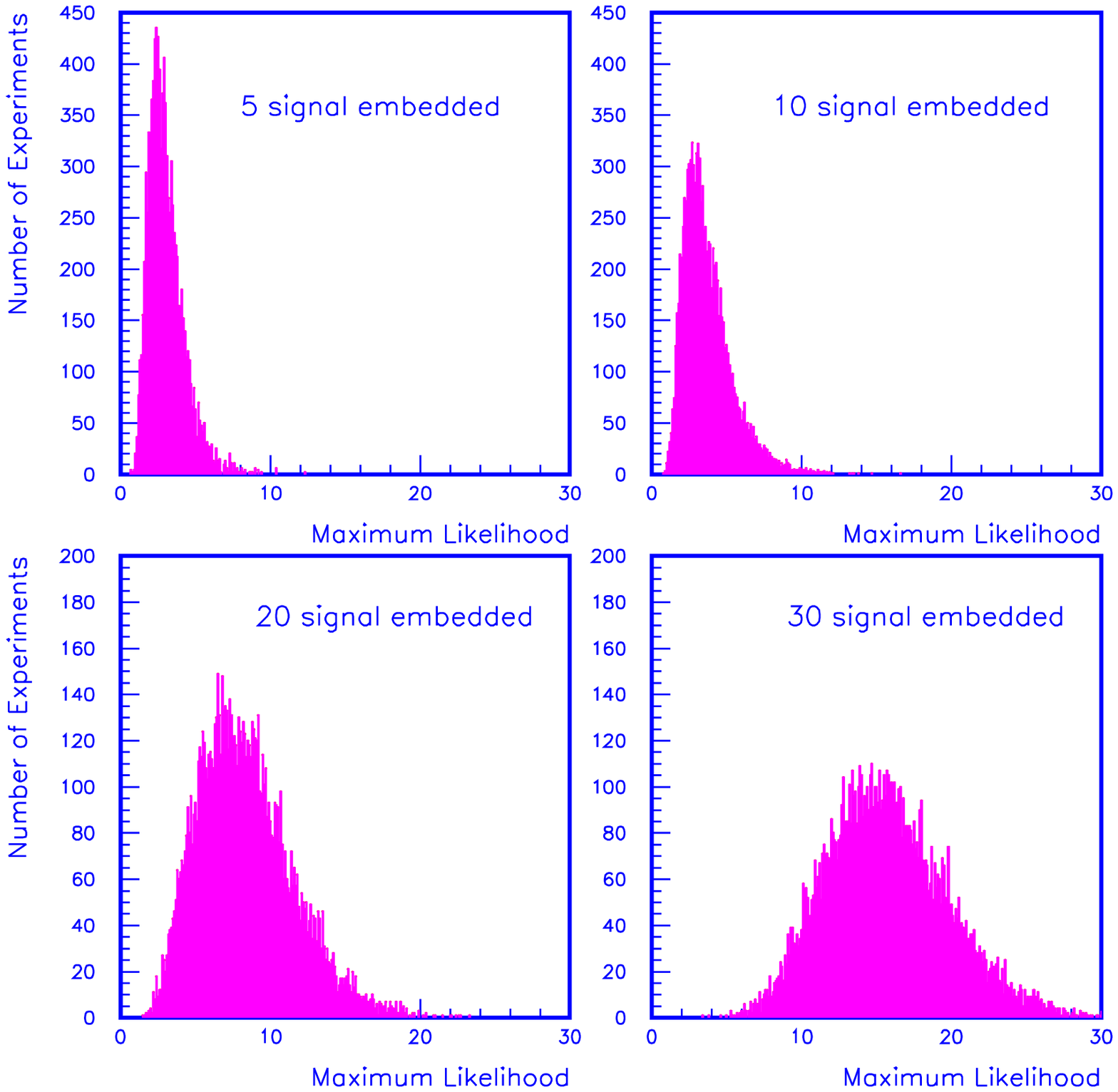}}}
\caption{The Maximum Likelihood output for MC experiments with 5, 10, 20, 30
         signal events embedded respectively.}
\label{fig:ML-signal}
\end{figure}

The work and results for 5, 10, 15, 20, 25, 30 and 35 signal events embedded with 
Gaussian means at 42.00, 46.00, 48.00, 49.00, 49.50, 49.75, 49.90 and 50.00 are 
shown in Ref.~\cite{liangthesis}.
The results for 10, 20, and 30 signal events embedded are shown in
Tables~\ref{result10},~\ref{result20} and ~\ref{result30}.
We can see that the number of signal embedded experiments successfully found
with a certain significance is much lower than what expected from $S/\sqrt{B}$
calculations.
This is a price we have to pay for not knowing the exact location of the signal. 
Furthermore, for the ``Sliding-Window'' method, the sensitivity strongly depends 
on the exact location of the embedded signal. If the step size is greater 
than 1, the embedded signals are totally missed for the worst-case scenarios. 
For step size of 1 or less, there are still significant differences in the 
sensitivities between the best-case and worse-case scenarios, depends on the 
step size of the ``Sliding-Window'' used to scan the kinematic range. 
In comparison, the new analysis method is independent of the exact location of
the Gaussian mean of the embedded signal events. 
This is because the new method scans 
the entire parameter space for the best fit to the entire experiment. 
The maximum  $S/\sqrt{B}$ from ``Sliding-Window'' approaches
with step sizes of 16, 4, 1, and 0.1 for the best-case scenario MC experiments 
each with 20 signal events embedded are shown in Figure~\ref{fig:SW-20-best}.
Similarly, 
the maximum  $S/\sqrt{B}$ from ``Sliding-Window'' approaches
with step sizes of 16, 4, 1, and 0.1 for the worst-case scenario MC experiments 
each with 20 signal events embedded are shown in Figure~\ref{fig:SW-20-worst}.
The maximum likelihood output of the best fits for MC experiments with 5, 10,
20 and 30 signal embedded are shown in Figure~\ref{fig:ML-signal}.
Compared to ``Sliding-Window'' approaches with a step size small enough not
to miss the worst-case scenarios, the sensitivity of the new analysis method
is significantly higher. This means that the new analysis approach is a
significantly better and more sensitive scheme to search for new physics signals
at the LHC than the current ``Sliding-Window'' method.

The analysis method described above performs a scan of the entire parameter 
space using unbinned maximum likelihood fits at every small interval of
the parameter space. It is very CPU-intensive. 
The 13.45 million background-only and 630,000 signal embedded MC experiments were 
generated and analyzed over several months with about 10 dual-CPU servers.

\section{Summary and Discussion}

We have examined the significance calculation and analysis methods 
in searching for an individual decay mode of a new physics signal  
at the LHC. Unlike the search for a physics signal with known 
location and shape, 
the significance calculation for new physics signals with 
unknown location or shape strongly depends on the details of the 
search scheme and the situation it applies to.
Using a signal with known shape but unknown location
as an example, we have demonstrated that the significance calculation 
using the current ``Sliding-Window'' method at the LHC is over-estimated. 
This is because we search for an excess of events over multiple 
narrow windows, but the significance is still calculated according 
to an individual narrow window.
The significance and sensitivity of the ``Sliding-Window'' method
strongly depends on the specifics of the method and the situation it
applies to, e.g. the step size of the ``Sliding-Window'' used to scan
the available kinematic range, the total available kinematic range
to search for the new physics signal, and the exact location of the 
new physics signal, etc. 

We describe general procedures for significance calculation and
comparing different search schemes. We have applied the procedures
and compared the current ``Sliding-Window'' approaches with a new 
analysis method. The proposed new analysis method uses maximum 
likelihood fits with floating parameters and scans the parameter 
space for the best fit to the entire sample.
We find the results of the new analysis method is independent of 
the location of the new physics signal and significantly more 
sensitive in searching for 
new physics signal than the current ``Sliding-Window'' approaches.

While the LHC experiments have great potential in
discovering many possible new physics signals, we need to be
extremely careful in evaluating the significance of an observation
from the real LHC data. Because possible new physics can show up in
many kinematic observables, over a very large kinematic range,
the fluctuation probability of background events will be much
higher.
For individual decay modes of new physics signals,
the expected significances in observing the new physics signal
will be much smaller than current 
expectations~\cite{atlas-tdr-15,atlasnotes,cmsnotes}. 
Combining independent decay modes of the same new physics
signal will be essential to establish the discovery of the new
physics signal. 
Significant observations of the same new particle in independent 
decay modes at consistent locations will be the most effective way 
to establish the discovery of this new particle.
Careful evaluation of the observation significance in each individual 
decay mode following the general procedures described in this paper
is the starting point, before we can evaluate the
significance of the observations of independent decay modes.

\section{Acknowledgment}

The authors would like to thank the members of the SMU HEP 
group for their encouragements and useful discussions. 
This work is supported by the U.S. Department of Energy under 
grant number DE-FG02-04ER41299.

\end{document}